\documentclass[runningheads]{llncs}
\usepackage[utf8]{inputenc}
\usepackage{fontenc}
\usepackage{amsmath,amssymb,amsfonts}
\usepackage{graphicx}
\usepackage{color}
\usepackage{subfigure}
\usepackage{fontspec}
\usepackage{balance}
\usepackage{textcomp}
\usepackage{cite}
\usepackage{hyperref}
\usepackage{multicol,lipsum}
\usepackage{algorithm}
\usepackage{tabularx}
\usepackage{caption} 
\usepackage{adjustbox}
\usepackage{booktabs}
\usepackage{algorithmic}
\usepackage{multirow}
\usepackage{hyperref}
\usepackage{mathtools}

\usepackage{bm}
\usepackage{bbding}
\usepackage{listings}
\usepackage{graphicx}
\usepackage{subcaption}
\usepackage{float}

\newcommand{\oomit}[1]{}

\begin{document}
\title{Comparative Analysis of Barrier-like Function Methods for Reach-Avoid Verification in Stochastic Discrete-Time Systems}
%
%
\author{Zhipeng Cao$^1$, Peixin Wang$^2$, Luke Ong$^3$, \DJ{}or\dj e \v{Z}ikeli\'c$^4$, Dominik Wagner$^3$, and Bai Xue$^1$}
%
%
\institute{1. KLSS and SKLCS, ISCAS, Beijing, China; School of Advanced Interdisciplinary Sciences, University of Chinese Academy of Sciences, Beijing, China \\
2. Software Engineering Institute, East China Normal University, China \\
3. College of Computing and Data Science, Nanyang Technological University, Singapore\\ 
4. Singapore Management University, Singapore, Singapore 
}
%
\maketitle              
\begin{abstract}
In this paper, we compare several representative barrier-like conditions from the literature for infinite-horizon reach-avoid verification of stochastic discrete-time systems. Our comparison examines both their theoretical properties and computational tractability, highlighting each condition’s strengths and limitations that affect applicability and conservativeness. Finally, we illustrate their practical performance through computational experiments using semidefinite programming (SDP) and counterexample-guided inductive synthesis (CEGIS).

\end{abstract}
\section{Introduction}
Reach-avoid verification ensures a system reaches target states while avoiding unsafe ones, and is fundamental for safety- and performance-critical applications such as autonomous driving, robotics, air traffic management, and systems biology \cite{summers2010verification,fisac2015reach,chen2016multiplayer,bajcsy2019efficient}. For example, a drone must reach its delivery location while avoiding obstacles; without reach-avoid guarantees, it may take risky paths. In deterministic systems, verification is usually qualitative, giving a binary outcome \cite{baier2008principles}, but for stochastic systems, probabilistic dynamics make strict guarantees too restrictive. Thus, quantitative verification is especially useful \cite{prajna2007framework,akametalu2014reachability}, which checks whether the probability of success exceeds a threshold.
A common approach to reach-avoid verification in stochastic systems uses barrier-like functions. Introduced in \cite{prajna2007framework} based on Ville’s inequality \cite{ville1939etude}, these functions initially provide probabilistic certificates that the system stays within a safe set with high probability. Unlike probabilistic model checking, which requires exploring all possible trajectories, barrier functions use analytical conditions to ensure safety more efficiently, even in high-dimensional or continuous spaces. Later, many variants have been developed for both finite- and infinite-time safety verification. Examples include $\epsilon$-repulsing supermartingales, $c$-martingales, and $k$-inductive barrier certificates \cite{chatterjee2017stochastic, steinhardt2012finite, jagtap2020formal, santoyo2021barrier, bak2018t, anand2022k, zhi2024unifying}. Compared with safety verification, reach-avoid verification for stochastic systems has received less attention. But interest in this topic has grown rapidly in the past five years, leading to several new barrier methods. For example, \cite{chatterjee2017stochastic} combines ranking supermartingales with stochastic invariants for analyzing termination in probabilistic programs—a setting closely related to reach-avoid verification in stochastic discrete-time systems. In \cite{xue2021reach,xue2024sufficient}, barrier-like functions were proposed by relaxing equality conditions that define the reach-avoid probability. Furthermore, additive and multiplicative reach-avoid supermartingales have been proposed in \cite{vzikelic2023learning, vzikelic2023compositional} for both verification and controller synthesis. 
Besides, computational strategies for synthesizing such functions range from linear and semidefinite programming \cite{xue2021reach} to more general constraint-solving techniques \cite{chatterjee2017stochastic,chatterjee2022sound} and neurosymbolic methods \cite{zhao2020synthesizing,abate2021fossil,vzikelic2023learning, vzikelic2023compositional,badings2025policy}, each offering distinct trade-offs between tractability and expressiveness. As these approaches continue to diversify, it becomes important to compare different barrier-like conditions in a systematic way—to clarify their theoretical relationships and assess how well they work on different types of systems. 




In this paper, we compare barrier-like conditions for infinite-horizon reach-avoid verification in stochastic discrete-time systems from both theoretical and computational perspectives. Our main contributions are: (1) a unified comparison of several key barrier-like conditions; (2) clarification of their theoretical relationships, including implications, equivalences, and relative strengths; (3) identification of key assumption such as robust invariance and strict supermartingale decrease that affect their applicability and conservativeness; and (4) computational experiments using SDP and CEGIS to demonstrate their practical performance.


\subsection*{Related Work}
Barrier certificates, inspired by Lyapunov functions in stability analysis, were first introduced for hybrid systems as a formal tool for safety verification \cite{prajna2004safety}. Like Lyapunov functions, the existence of a barrier function can guarantee that safety or reachability specifications are satisfied \cite{prajna2007convex}. Later work extended and improved barrier functions and expanded their applications \cite{ames2019control,taylor2020learning,xue2023reacho}. In the stochastic setting, safety verification over the infinite time horizon via barrier certificates was introduced alongside its deterministic counterpart in \cite{prajna2007framework}. Utilizing Ville’s Inequality \cite{ville1939etude}, \cite{prajna2007framework} constructed a non-negative barrier function and provided a sufficient condition for upper bounding the probability of eventually entering an unsafe region while remaining within a state constraint set for continuous-time systems. More recently, a new barrier function, constructed by relaxing a set of equations, was proposed in \cite{xue2021reach} for computing $p$-reach-avoid sets in stochastic discrete-time systems, which include initial states from which the system can reach the target set while staying in the safe set along the way, with probability at least $p$. 
This approach was subsequently extended to continuous-time systems in \cite{xue2024}. These barrier functions were further generalized in \cite{yu2023safe} to determine the lower and upper bounds of the safety probability over an infinite time horizon. Additionally, by relaxing Bellman equations, \cite{xue2024sufficient} provided necessary and sufficient conditions to bound safety and reach-avoid probabilities for stochastic discrete-time systems over an infinite horizon. Assuming the system evolves within a robust invariant set, \cite{vzikelic2023learning,vzikelic2023compositional} introduced two new barrier certificates, i.e., additive and multiplicative reach-avoid supermartingales, to ensure reach-avoid specifications and to support reach-avoid controller design. Barrier certificates have also been extended to infinite-time probabilistic program analysis to verify properties like almost-sure termination, probabilistic termination, assertion violations, and reachability \cite{chakarov2013probabilistic,mciver2017new,moosbrugger2021automated,kenyon2021supermartingales,chatterjee2017stochastic,chatterjee2022sound,majumdar2025sound,wang2021quantitative,takisaka2021ranking}. Probabilistic termination analysis \cite{chatterjee2017stochastic,chatterjee2022sound,majumdar2025sound} is closely related to classical stochastic system analysis. Finding a lower bound on termination probability is like reach-avoid analysis, measuring the chance a program reaches terminal states while staying within a stochastic invariant. Conversely, finding an upper bound is similar to safety analysis, representing the chance a program leaves the invariant before reaching a terminal state.

Finite-time verification has also been extensively studied. Building on \cite{kushner1967stochastic}, \cite{steinhardt2012finite} introduced $c$-martingales for stochastic differential equations, allowing controlled growth in a certificate’s expectation to bound finite-horizon safety violations. This idea was later extended to discrete-time temporal-logic verification under invariant-set assumptions \cite{jagtap2018temporal,jagtap2020formal}, and was further developed for finite-time safety and controller synthesis via semidefinite programming \cite{santoyo2021barrier}. More recently, barrier functions have been applied to bound probabilistic safety over finite and infinite horizons \cite{zhi2024unifying}, with invariant set assumptions later relaxed \cite{xue2023new,xue2024finite,chen2025construction,xue2025refined}. Extending the concept further, barrier certificates have also been applied to analyze probabilistic programs with bounded time horizons (e.g., \cite{chatterjee2017stochastic,kura2019tail,chatterjee2024quantitative}). For instance, \cite{chatterjee2017stochastic} (see Lemma 3) presented a sufficient condition based on an $\epsilon$-repulsing supermartingale supported by a pure invariant to derive upper bounds on the probability that programs reach a specific set exactly at a given step. When programs terminate almost surely, \cite{kura2019tail} proposed sufficient conditions to establish lower bounds on termination within bounded time horizons. Similarly, \cite{wang2021central} developed bounds for the tail probability problem, which can also estimate the likelihood of program termination within bounded horizons. Recently, using stochastic invariants \cite{chatterjee2017stochastic}, \cite{chatterjee2024quantitative} studied tail bounds for programs that may not terminate almost surely. Their method yields lower bounds on the probability that a program terminates within a bounded time.

The paper is organized as follows. Section~\ref{sec:pre} introduces stochastic discrete-time systems and the infinite-horizon reach-avoid problem. Section~\ref{sec:CA} compares several existing barrier-like conditions. Section~\ref{sec:ex} presents numerical examples using SDP and CEGIS. Section~\ref{sec:con} concludes the paper.

The following notations are used throughout this paper: $\mathbb{R}$: real numbers; $\mathbb{R}_{\ge 0}$: non-negative real numbers; $\mathbb{N}$: nonnegative integers; $\mathbb{N}_{\le k} = \{n \in \mathbb{N} \mid n \le k\}$, $\mathbb{N}_{\ge k} = \{n \in \mathbb{N} \mid n \ge k\}$.  For sets $\Delta_1$ and $\Delta_2$, $\Delta_1\setminus \Delta_2 = \{\bm{x}\in \Delta_1 \mid \bm{x}\notin \Delta_2\}$;
$1_A(\bm{x})$ is the indicator of set $A$, equal to 1 if $\bm{x}\in A$ and 0 otherwise.
For any set $\Delta$, $\Delta^\infty$ denotes the set of all infinite sequences over $\Delta$.
\section{Preliminaries}
\label{sec:pre}
In this section, we introduce stochastic discrete-time systems and the reach-avoid verification problem. 

\subsection{Problem Statement}
\label{sub:ps}
This paper considers stochastic discrete-time systems that are modeled by stochastic difference equations of the following form:
\begin{equation}
\label{system}
\begin{split}
\bm{x}(l+1)=\bm{f}(\bm{x}(l),\bm{\theta}(l)), \forall l\in \mathbb{N},
\end{split}
\end{equation}
where $\bm{x}(l) \in \mathbb{R}^n$ is the state at time $l$ and $\bm{\theta}(l)\in \Theta$ with $\Theta \subseteq \mathbb{R}^m$ is the stochastic disturbance at time $l$. The disturbances $\bm{\theta}(0), \bm{\theta}(1), \ldots$ are i.i.d. (independent and identically distributed) random variables on a probability space $(\Theta,\mathcal{F},\mathbb{P}_{\bm{\theta}})$, and take values in $\Theta$ with the probability distribution: for any measurable set $B\subseteq \Theta$, ${\rm Prob}(\bm{\theta}(l)\in B)=\mathbb{P}_{\bm{\theta}}(B), \forall l\in \mathbb{N}$. The expectation is denoted as $\mathbb{E}_{\bm{\theta}}[\cdot]$.

Before defining the trajectory of system \eqref{system}, we define a disturbance signal.
\begin{definition}

A disturbance signal $\pi$ is an element of the set $\Theta^\infty$ (i.e., an infinite sequence $\pi = (\bm{\theta}(0), \bm{\theta}(1), \dots)$ where each $\bm{\theta}(i) \in \Theta$). The set $\Theta^\infty$ is endowed with the product topology and the corresponding Borel $\sigma$-algebra $\mathcal{B}(\Theta^\infty)$. The probability measure on this space is given by $\mathbb{P}_{\pi} := \mathbb{P}_{\bm{\theta}}^{\infty}$ (the infinite product measure of $\mathbb{P}_{\bm{\theta}}$), with corresponding expectation operator $\mathbb{E}_{\pi}[\cdot]$.
\end{definition}


A disturbance signal $\pi$ together with an initial state $\bm{x}_0\in \mathbb{R}^n$ induces a unique discrete-time trajectory as follows.
\begin{definition}
Given a disturbance signal $\pi$ and an initial state $\bm{x}_0\in \mathbb{R}^n$, a trajectory of system \eqref{system} is denoted as  $\bm{\phi}_{\pi}^{\bm{x}_0}(\cdot)\colon\mathbb{N}\rightarrow \mathbb{R}^n$ with $\bm{\phi}_{\pi}^{\bm{x}_0}(0)=\bm{x}_0$ and $\bm{\phi}_{\pi}^{\bm{x}_0}(l+1)=\bm{f}(\bm{\phi}_{\pi}^{\bm{x}_0}(l),\bm{\theta}(l)), \forall l\in \mathbb{N}$.
\end{definition}

The infinite-time reach-avoid verification  problem is defined below. 

\begin{definition}[Reach-avoid Verification]
\label{reach-avoid}
Given a safe set $\mathcal{X}\subseteq \mathbb{R}^n$, an initial set $\mathcal{X}_0\subseteq \mathcal{X}\setminus \mathcal{T}$, a target set $\mathcal{T} \subseteq \mathcal{X}$, and a probability threshold $p\in (0,1]$, the reach-avoid verification aims to determine whether the reach-avoid probability $\mathbb{P}_{\pi}(RA_{\bm{x}_0})$, which denotes the probability that the system \eqref{system}, starting from any state $\bm{x}_0\in \mathcal{X}_0$, will reach the target set $\mathcal{T}$ eventually while staying within the safe set $\mathcal{X}$ before hitting the target set, is greater than or equal to $p$, i.e., $\mathbb{P}_{\pi}(RA_{\bm{x}_0})\geq p, \forall \bm{x}_0\in \mathcal{X}_0$, where $RA_{\bm{x}_0}=\{\pi \mid \exists k\in \mathbb{N}. \bm{\phi}_{\pi}^{\bm{x}_0}(k)\in \mathcal{T} \wedge \forall i\in \mathbb{N}_{\leq k}.  \bm{\phi}_{\pi}^{\bm{x}_0}(i)\in \mathcal{X}\}$.
\end{definition}

\section{Comparative Analysis}
\label{sec:CA}
This section compares several barrier-like conditions for reach-avoid verification. We present them in order of their strength in certifying reach-avoid properties and discuss their expressiveness, assumptions, and practical use.

\subsection{Barrier-like Conditions I}
We analyze a barrier-like condition combining a stochastic invariant with an $\epsilon$-ranking supermartingale. This condition and its variants have been applied to reach–avoid verification (e.g., implicitly in \cite{abate2025quantitative}) and to analyzing termination of probabilistic programs (e.g., \cite{chatterjee2017stochastic, chatterjee2022sound, majumdar2025sound}). In termination analysis, $\mathcal{T}$ typically represents terminal states or an invariant set that traps all future transitions.

The condition requires two functions: a supermartingale barrier function $h_1 : \mathbb{R}^n \rightarrow \mathbb{R}_{\geq 0}$ and an $\epsilon$-ranking supermartingale $h_2 : \mathbb{R}^n \rightarrow \mathbb{R}_{\geq 0}$, where $\epsilon>0$ is a constant. The function $h_1 : \mathbb{R}^n\rightarrow \mathbb{R}_{\geq 0}$ satisfies three conditions: 1)  it is less than or equal to $1-p$ over the initial set $\mathcal{X}_0$; 2) $h_1(\bm{x})\geq 1$ for all $\bm{x}\in \mathbb{R}^n\setminus \mathcal{X}$; and 3) within the safe set $\mathcal{X}$, its expected value must not increase at each time step. 
 $h_2(\bm{x})$ is required to be non-negative over $\mathbb{R}^n$ and its expected value must decrease by at least  $\epsilon$ at each step on the set $\mathcal{X}\setminus \mathcal{T}$. 
\begin{theorem}
\label{first}
    If there exist two functions $h_1 :\mathbb{R}^n \rightarrow \mathbb{R}$ and $h_2 : \mathbb{R}^n \rightarrow \mathbb{R}$, and a positive value $\epsilon>0$, such that the following constraints are satisfied:
    \begin{equation}
    \tag{BC1}
    \label{first_barrier}
        \begin{cases}
            h_1(\bm{x})\leq 1-p, &\forall \bm{x}\in \mathcal{X}_0,\\
            \mathbb{E}_{\bm{\theta}}[h_1(\bm{f}(\bm{x},\bm{\theta}))\leq h_1(\bm{x}), & \forall \bm{x} \in \mathcal{X},\\
            h_1(\bm{x})\geq 1, &\forall \bm{x}\in 
            \mathbb{R}^n\setminus \mathcal{X},\\
            h_1(\bm{x})\geq 0, &\forall \bm{x}\in \mathcal{X},\\
            h_2(\bm{x})\geq 0, &\forall \bm{x}\in \mathbb{R}^n,\\
            \mathbb{E}_{\bm{\theta}}[h_2(\bm{f}(\bm{x},\bm{\theta}))]-h_2(\bm{x})\leq -\epsilon, & \forall \bm{x} \in \mathcal{X}\setminus \mathcal{T},
        \end{cases}
    \end{equation}
then for every $\bm{x}_0\in \mathcal{X}_0$, $\mathbb{P}_{\pi}(RA_{\bm{x}_0})\geq 1-h_1(\bm{x}_0)\geq p$.
\end{theorem}

\textbf{Theoretical Analysis}: Condition \eqref{first_barrier} involves two functions, $h_1$ and $h_2$, and resembles a barrier–Lyapunov condition, which enforces both safety (state constraints) and convergence \cite{ames2016control,wang2018permissive}. The key difference is that~\eqref{first_barrier} focuses on reaching a target set rather than stabilizing the system to a single equilibrium.

The function $h_1$ ensures that, starting from any $\bm{x}_0\in \mathcal{X}_0$, the probability of staying in the safe set $\mathcal{X}$ at all times is at least $1-h_1(\bm{x}_0)$ (Corollary \ref{first_coro} below). The function $h_2$ ensures, via the Foster–Lyapunov drift condition, that conditioned on remaining in $\mathcal{X}$, trajectories starting from $\bm{x}_0$ reach the target $\mathcal{T}$ with probability one. Ideally, $1-h_1(\bm{x}_0)>0$ when the reach-avoid probability is positive, but this can be overly restrictive for systems in which all trajectories eventually leave the safe set almost surely—for example, a system $\bm{x}(k+1)=\bm{g}(\bm{x}(k))+\bm{\theta}(k)$ with Gaussian disturbances $\bm{\theta}(k)$ and a bounded safe set $\mathcal{X}$. For this system, $1 - h_1(\bm{x}_0)$ must be zero, since trajectories starting from any state in $\mathcal{X}$ will almost surely leave this set. However, even though no $h_1$ can satisfy $h_1(\bm{x}_0)<1$ in this case, some trajectories may still reach $\mathcal{T}$ before exiting, so the overall reach-avoid probability can remain positive. While the constraint \eqref{first_barrier} is generally strict, it is suitable when the target set is invariant, contains an invariant subset, or consists of terminal states. Consequently, this constraint is commonly used in probabilistic program termination analysis \cite{chatterjee2017stochastic,chatterjee2022sound,majumdar2025sound} and in reach-avoid-stay problems \cite{meng2022smooth,meng2022sufficient,neustroev2025neural}.  On the other hand, if $h_2:\mathbb{R}^n \to \mathbb{R}$ is bounded, we can further conclude that, for almost all trajectories that remain in $\mathcal{X}$, the process not only reaches $\mathcal{T}$ in finite time, but also satisfies one of the following: either it eventually stays in $\mathcal{T}$ forever, or it returns to $\mathcal{T}$ infinitely many times.

\begin{corollary}
\label{first_coro}
If there exist two functions $h_1, h_2 : \mathbb{R}^n \rightarrow \mathbb{R}$ and a positive value $\epsilon>0$ satisfying the constraint \eqref{first_barrier}, then $\mathbb{P}_{\pi}(\forall i\in \mathbb{N}. \bm{\phi}_{\pi}^{\bm{x}_0}\in \mathcal{X})\geq 1-h_1(\bm{x}_0)$, $\forall   \bm{x}_0\in \mathcal{X}$. Furthermore, when $p>0$ and $h_2 : \mathbb{R}^n \rightarrow \mathbb{R}$ is bounded, we have $\mathbb{P}_{\pi}(\sum_{i=1}^{\infty}1_{\mathcal{T}}(\bm{\phi}_{\pi}^{\bm{x}_0}(i))=\infty\mid \forall i\in \mathbb{N}. \bm{\phi}_{\pi}^{\bm{x}_0}(i)\in \mathcal{X})=1$.
\end{corollary}

The proof of Corollary \ref{first_coro} is shown in Appendix. If there exist functions $h_1, h_2$ and a constant $\epsilon>0$ satisfying \eqref{first_barrier}, the system starting in $\mathcal{X}\setminus \mathcal{T}$ will almost surely leave this set in finite time. In this context, 
a related necessary and sufficient condition was later proposed in \cite{xue2024sufficient} by relaxing a Bellman equation.

\begin{proposition}[Theorem 2, \cite{xue2024sufficient}]
\label{as}
Assume $\mathbb{P}_{\pi}(\forall i\in \mathbb{N}.\bm{\phi}_{\pi}^{\bm{x}_0}(i)\in \mathcal{X}\setminus \mathcal{T})=0, \forall \bm{x}_0\in \mathcal{X}\setminus \mathcal{T}$. There exists a function $v : \mathbb{R}^n \rightarrow \mathbb{R}$ satisfying 
\begin{equation}
    \label{as_barrier}
    \begin{cases}
        v(\bm{x})\geq p, &\forall \bm{x}\in \mathcal{X}_0,\\
        v(\bm{x}) \leq \mathbb{E}_{\bm{\theta}}[v(\bm{f}(\bm{x},\bm{\theta}))], &\forall \bm{x}\in \mathcal{X}\setminus \mathcal{T},\\
        v(\bm{x})\leq 1, &\forall \bm{x}\in \mathcal{T}, \\
        v(\bm{x}) \leq 0, &\forall \bm{x} \in \mathbb{R}^n\setminus \mathcal{X},
    \end{cases}
\end{equation}
if and only if $\mathbb{P}_{\pi}(RA_{\bm{x}_0}) \geq p, \forall \bm{x}_0\in \mathcal{X}_0$.  
 \end{proposition}


Unlike \eqref{first_barrier} in Theorem \ref{first}, which requires two functions, the condition \eqref{as_barrier} involves only a single function. In addition, we observe that if $h_1$ and $h_2$ satisfy \eqref{first_barrier}, then $v(\bm{x}) := 1 - h_1(\bm{x})$ also satisfies \eqref{as_barrier}. Since \eqref{as_barrier} provides a necessary and sufficient condition when $\mathbb{P}_{\pi}(\forall i\in \mathbb{N}. \bm{\phi}_{\pi}^{\bm{x}_0}(i) \in \mathcal{X}\setminus \mathcal{T}) = 0$ for all $\bm{x}_0 \in \mathcal{X}\setminus \mathcal{T}$, it can be used in two ways: (a) to improve the lower bound $p$ of $\mathbb{P}_{\pi}(RA_{\bm{x}_0})$ when a solution to \eqref{first_barrier} exists, or (b) as an alternative for reach-avoid verification if no solution to \eqref{first_barrier} is found.

\textbf{Computational Tractability}: Condition \eqref{first_barrier} is convex in $h_1(\bm{x})$ and $h_2(\bm{x})$, meaning any convex combination of two solutions is also a solution. This convexity has both theoretical and practical benefits. For polynomial systems, searching for suitable functions can be formulated as a convex optimization problem using tools like SOSTOOLS \cite{prajna2002introducing} and advanced algorithms \cite{parrilo2000structured,lasserre2001global}. Convexity also improves robustness, making the solution less sensitive to data perturbations, or numerical errors, which is valuable in practical computations. In addition, the condition \eqref{as_barrier} is also convex in $h(\bm{x})$. 

\subsection{Barrier-like Conditions II}
This section analyzes two barrier-like conditions from \cite{vzikelic2023learning, vzikelic2023compositional}, whose corresponding functions are called additive and multiplicative reach-avoid supermartingales.

The first barrier-like condition, proposed in \cite{vzikelic2023learning}, addresses reach-avoid verification and controller synthesis for systems evolving within a prescribed invariant set $\Omega$. It uses a continuous function $V:\Omega\to\mathbb{R}$, called an additive-reach-avoid supermartingale(ARAS), which satisfies: (1) $V(\bm{x})\ge 0$ on $\Omega$, (2) $V(\bm{x})\le 1$ on the initial set $\mathcal{X}_0$, (3) $V(\bm{x})\ge \frac{1}{1-p}$ outside the safe set, and (4) its expected value decreases by at least $\epsilon>0$ in one step for all $\bm{x}$ in $\{\Omega\setminus \mathcal{T} \mid V(\bm{x})\leq  \frac{1}{1-p}\}$.

\begin{theorem} [\cite{vzikelic2023learning}]
\label{second}
Let $\Omega$ be a robust invariant for system \eqref{system}, (i.e., $\bm{f}(\bm{x},\bm{\theta}) \in \Omega$ for any $\bm{x}\in \Omega$, $\bm{\theta}\in \Theta$), $\mathcal{T}\subset \Omega$, $\mathcal{X}\subset \Omega$, $\mathcal{X}_0\subseteq \mathcal{X}$, and $p\in [0,1)$.
If there exist a continuous function $V : \Omega\rightarrow \mathbb{R}$ and a positive value $\epsilon>0$ satisfying 
\begin{equation}
\tag{BC2}
\label{second_barrier}
    \begin{cases}
        V(\bm{x}) \geq 0, &\forall \bm{x}\in \Omega,\\
        V(\bm{x}) \leq 1, & \forall \bm{x}\in \mathcal{X}_0,\\
        V(\bm{x}) \geq \frac{1}{1-p}, &\forall \bm{x}\in \Omega\setminus \mathcal{X},\\
        \mathbb{E}_{\bm{\theta}}[V(\bm{f}(\bm{x},\bm{\theta}))]-V(\bm{x})\leq -\epsilon, &\forall \bm{x}\in \{\Omega\setminus \mathcal{T}\mid V(\bm{x})\leq \frac{1}{1-p}\}, 
    \end{cases}
\end{equation}
then, for every $\bm{x}_0\in \mathcal{X}_0$, $\mathbb{P}_{\pi}(RA_{\bm{x}_0})\geq p$. 
\end{theorem}

The function $V : \Omega\rightarrow \mathbb{R}$ in Theorem \ref{second} is called an $\epsilon$-ARAS in \cite{vzikelic2023learning}, and an $(\epsilon,\lambda)$-ARAS  in~\cite{vzikelic2023compositional}, where $\lambda = \frac{1}{1-p}$.

The second barrier-like condition was proposed in \cite{vzikelic2023compositional}. 
It uses a continuous function $V:\Omega\to\mathbb{R}$, called a multiplicative-reach-avoid supermartingale(MRAS), which satisfies: (1) $V(\bm{x})\ge 0$ on $\Omega$ (a robust invariant set), (2) $V(\bm{x})\ge \delta>0$ on $\Omega\setminus \mathcal{T}$, (3) $V(\bm{x})\le 1$ on $\mathcal{X}_0$, (4) $V(\bm{x})\ge \lambda'>1$ outside the safe set, and (5) its $\gamma$-scaled value dominates the expected next-step value for all $\bm{x}$ in $\{\Omega\setminus \mathcal{T} \mid V(\bm{x})\le \lambda'\}$. If such a function exists, the reach-avoid probability from each initial state in $\mathcal{X}_0$ is at least $1-\frac{1}{\lambda'}$.
 
\begin{theorem}[Theorem 1, \cite{vzikelic2023compositional}]
\label{third_one_barrier}
Let $\Omega$ be a robust invariant for system \eqref{system}, 
$\mathcal{T}\subset \Omega$, $\mathcal{X}\subset \Omega$, $\mathcal{X}_0\subseteq \mathcal{X}$, $\gamma\in (0,1)$, $\delta>0$, $\lambda'>1$, and $1-\frac{1}{\lambda'}\in [0,1)$. 
If there exists a continuous function $V: \Omega \rightarrow \mathbb{R}$ satisfying 
\begin{equation}
\tag{BC3}
\label{mul_barrier}
    \begin{cases}
        V(\bm{x})\geq 0, &\forall \bm{x} \in \Omega,\\
        V(\bm{x}) \geq \delta, & \forall \bm{x}\in \Omega\setminus \mathcal{T},\\
        V(\bm{x}) \leq 1, &\forall \bm{x} \in \mathcal{X}_0,\\
        V(\bm{x}) \geq \lambda', & \forall \bm{x}\in \Omega\setminus \mathcal{X},\\
        \gamma V(\bm{x}) \geq \mathbb{E}_{\bm{\theta}}[V(\bm{f}(\bm{x},\bm{\theta}))], &\forall \bm{x} \in \{\Omega\setminus \mathcal{T}\mid V(\bm{x})\leq \lambda'\},
    \end{cases}
\end{equation}
then, $\mathbb{P}_{\pi}(RA_{\bm{x}_0})\geq 1-\frac{1}{\lambda'}$, $\forall \bm{x}\in \mathcal{X}_0$.
\end{theorem}

   The function $V: \Omega \rightarrow \mathbb{R}$ in Theorem \ref{third_one_barrier} is called a $(\gamma,\delta,\lambda')$-MRAS in \cite{vzikelic2023compositional}. 
    The relationship between MRAS and ARAS is presented in \cite{vzikelic2023compositional}:
\begin{enumerate}
    \item if $V: \Omega \rightarrow \mathbb{R}$ is an $(\epsilon,\lambda)$-ARAS, then it is also a $(\frac{\lambda-\epsilon}{\lambda},\min\{\epsilon,\lambda\},\lambda)$-MRAS.
\item if $V: \Omega \rightarrow \mathbb{R}$ is an $(\gamma,\delta,\lambda)$-MRAS, then it is also a $((1-\gamma)\delta,\lambda)$-ARAS.
\end{enumerate}

\textbf{Theoretical Analysis:} 
Unlike condition \eqref{first_barrier} in Theorem \ref{first}, which uses two functions,  \eqref{second_barrier} in Theorem \ref{second} requires only one. This is a simple but important difference. Below, we provide a detailed analysis of condition \eqref{second_barrier}:

\noindent 1.  \textit{Domain Restriction and Robust Invariance Assumption:} Theorem \ref{second} assumes that $\Omega$ is a robust invariant set for the system \eqref{system}. This assumption has been widely adopted in the formal verification of stochastic discrete-time systems (e.g., \cite{chakarov2016deductive,jagtap2018temporal,jagtap2020formal,santoyo2021barrier,mazouz2022safety,salamati2024data,zhi2024robustness,zhi2024unifying,henzinger2025supermartingale,badings2025policy,feng2025runtime}) and probabilistic programs (e.g.,\cite{chatterjee2017stochastic,chatterjee2022sound, majumdar2025sound}). However, the limitations and potential conservativeness of this assumption on formal verification have been critically examined in \cite{xue2023reachability,yu2023safe,xue2024finite}. When $\Omega \neq \mathbb{R}^n$, this requirement can be stringent because many systems do not admit such a robust invariant set $\Omega$ \cite{khalil2002nonlinear}. If $\Omega$ is not a robust invariant set, the conclusions in Theorem \ref{second} and \ref{third_one_barrier} do not hold: the expectation $\mathbb{E}_{\bm{\theta}}[V(\bm{f}(\bm{x},\bm{\theta}))]$ in \eqref{second_barrier}/\eqref{mul_barrier} is ill-defined, because $V$ is defined only on $\Omega$, while $\bm{f}(\bm{x},\bm{\theta})$ may lie outside $\Omega$ with positive probability. 
In such cases, additional constraints (e.g., $V(\bm{x})\geq \frac{1}{1-p}$, $\forall \bm{x}\in \mathbb{R}^n\setminus \Omega$ in \eqref{second_barrier} and $V(\bm{x})\geq \lambda', \forall \bm{x}\in \mathbb{R}^n\setminus \Omega$ in \eqref{mul_barrier}) should be imposed to ensure that the expectation is \textit{well defined} 
and further guarantee the reach–avoid property. A similar discussion was presented in Remark 3 in \cite{xue2024finite}.
    
\noindent 2. \textit{Expectation Decrease Condition and Its Domain:} Similar to  $h_2(\bm{x})$ in Theorem \ref{first}, $V(\bm{x})$ in Theorem \ref{second} satisfies a strict expectation decrease condition, i.e.,
     $\mathbb{E}_{\bm{\theta}}[V(\bm{f}(\bm{x},\bm{\theta}))]-V(\bm{x})\leq -\epsilon$. However, unlike Theorem \ref{first}, this condition is enforced over the restricted set $\{\Omega\setminus \mathcal{T}\mid V(\bm{x})\leq \frac{1}{1-p}\}$ rather than the broader domain  $\mathcal{X}\setminus \mathcal{T}$. Constituently, this single condition cannot guarantee the existence of a positive measure of trajectories that remain within $\{\Omega\setminus \mathcal{T}\mid V(\bm{x})\leq \frac{1}{1-p}\}$ for all time, nor can it ensure that almost all trajectories reaching $\mathcal{T}$ while staying within $\{\bm{x}\in \Omega\mid V(\bm{x})\leq \frac{1}{1-p}\}$ will revisit $\mathcal{T}$ infinitely often when $V(\bm{x})$ is bounded. Trajectories may exit the set $\{\bm{x}\in \Omega\mid V(\bm{x})\leq \frac{1}{1-p}\}$ after reaching $\mathcal{T}$.
     
    \textbf{Computational Tractability}: 
    Both conditions \eqref{second_barrier} and \eqref{mul_barrier} are nonlinear in $V(\bm{x})$, because they require $\mathbb{E}_{\bm{\theta}}[V(\bm{f}(\bm{x},\bm{\theta}))]-V(\bm{x})\le -\epsilon$ and
$\gamma V(\bm{x}) \ge \mathbb{E}_{\bm{\theta}}[V(\bm{f}(\bm{x},\bm{\theta}))]$ to hold only on subsets of the state space defined by $V(\bm{x})$, namely
$\{\bm{x}\in \Omega\setminus \mathcal{T}\mid V(\bm{x})\le \frac{1}{1-p}\}$ and $\{\bm{x}\in \Omega\setminus \mathcal{T}\mid V(\bm{x})\le \lambda'\}$, respectively.

\subsection{Barrier-like Conditions III}
This section presents a barrier-like condition, originally proposed in \cite{xue2024sufficient}, which is obtained by relaxing a Bellman equation. A version of this condition for uncertain deterministic discrete-time systems is presented in \cite{zhao2022inner}.


This condition requires a function $h$ and a scalar $\lambda \in (0,1)$ satisfying: (1) $h \ge p$ on the initial set $\mathcal{X}_0$, (2) $h \le 0$ outside the safe set, (3) $h \le 1$ on the target set $\mathcal{T}$, and (4) the $\lambda$-scaled expected value of $h$ at the next step is at least its current value for $\bm{x} \in \mathcal{X} \setminus \mathcal{T}$. If such $h$ and $\lambda$ exist, the reach-avoid probability for any $\bm{x}_0 \in \mathcal{X}_0$ is at least $p$.

\begin{theorem}[\cite{xue2024sufficient}]
\label{third}
If there exist a function $h : \mathbb{R}^n \rightarrow \mathbb{R}$, which is bounded over $\mathcal{X}$, and a positive value $\lambda \in (0,1)$, satisfying 
\begin{equation}
\tag{BC4}
\label{third_barrier}
    \begin{cases}
        h(\bm{x}) \geq p, &\forall \bm{x}\in \mathcal{X}_0,\\
        h(\bm{x}) \leq 0, &\forall \bm{x}\in \mathbb{R}^n\setminus \mathcal{X},\\
        h(\bm{x}) \leq 1, & \forall \bm{x}\in \mathcal{T},\\
        h(\bm{x})\leq \lambda\mathbb{E}_{\bm{\theta}}[h(\bm{f}(\bm{x},\bm{\theta}))], &\forall \bm{x}\in \mathcal{X}\setminus \mathcal{T}, 
    \end{cases}
\end{equation}
then $\mathbb{P}_{\pi}(RA_{\bm{x}_0})\geq p$, $\forall \bm{x}_0\in \mathcal{X}_0$.
\end{theorem}

\textbf{Theoretical Analysis:} As shown in \cite{xue2024sufficient}, for each initial state $\bm{x}_0\in \mathcal{X}_0$, if the specified threshold $p$ is strictly less than $\mathbb{P}_{\pi}(RA_{\bm{x}_0})$-that is, $p<\mathbb{P}_{\pi}(RA_{\bm{x}_0})$-then there definitely exist a function $h : \mathbb{R}^n \rightarrow \mathbb{R}$, bounded over $\mathcal{X}$, and a scalar value $\lambda_{\bm{x}_0} \in (0,1)$, such that the following conditions hold:
\begin{equation}
\label{third_barrier0}
    \begin{cases}
        h(\bm{x}_0) \geq  p,\\
        h(\bm{x}) \leq 0, &\forall \bm{x}\in \mathbb{R}^n\setminus \mathcal{X},\\
        h(\bm{x}) \leq 1, & \forall \bm{x}\in \mathcal{T},\\
        h(\bm{x})\leq \lambda_{\bm{x}_0}\mathbb{E}_{\bm{\theta}}[h(\bm{f}(\bm{x},\bm{\theta}))], &\forall \bm{x}\in \mathcal{X}\setminus \mathcal{T}.
    \end{cases}
\end{equation}
When considering a continuous set of initial states rather than a single state, it remains an open question whether there exist a function $h : \mathbb{R}^n \rightarrow \mathbb{R}$, bounded over $\mathcal{X}$, and a scalar value $\lambda \in (0,1)$, that satisfy \eqref{third_barrier} uniformly for all $\bm{x}_0\in \mathcal{X}_0$, even in cases where $\sup_{\bm{x}_0\in \mathcal{X}_0}\mathbb{P}_{\pi}(RA_{\bm{x}_0})>p$.

On the other hand, under the same assumption as Theorem~\ref{second} that the system evolves within the robust invariant set $\Omega$, the following holds: if a function $V$ and $\epsilon>0$ satisfy condition \eqref{second_barrier}, then one can construct a function $h$ and a scalar $\lambda\in(0,1)$ that satisfy condition \eqref{third0_barrier} in Corollary~\ref{third_cor}, which is constructed by combining \eqref{third_barrier} and the function-dependent regions in \eqref{second_barrier}/\eqref{mul_barrier}. This means the feasible set of condition~\eqref{third0_barrier} is at least as large as that of \eqref{second_barrier}, making it potentially more powerful for reach-avoid verification. Similarly, if $V$, $\gamma\in(0,1)$, $\delta>0$, and $\lambda'>1$ satisfy condition~\eqref{mul_barrier}, then a function $h$ and scalar $\lambda\in(0,1)$ exist such that \eqref{third0_barrier} holds.

\begin{corollary}
\label{third_cor}
Let $\Omega$ be a robust invariant for system \eqref{system}, (i.e., $\bm{f}(\bm{x},\bm{\theta}) \in \Omega$ for any $\bm{x}\in \Omega$, $\bm{\theta}\in \Theta$), $\mathcal{T}\subset \Omega$, $\mathcal{X}\subset \Omega$, $\mathcal{X}_0\subseteq \mathcal{X}$, and $p\in [0,1)$ be a probability threshold. If there exist $h : \Omega\rightarrow \mathbb{R}$ and  $\lambda \in (0,1)$ satisfying 
\begin{equation}
\label{third0_barrier}
    \begin{cases}
        h(\bm{x}) \geq p, &\forall \bm{x}\in \mathcal{X}_0,\\
        h(\bm{x}) \leq 0, &\forall \bm{x}\in \Omega\setminus \mathcal{X},\\
        h(\bm{x}) \leq 1, & \forall \bm{x}\in \Omega,\\
        h(\bm{x})\leq \lambda \mathbb{E}_{\bm{\theta}}[h(\bm{f}(\bm{x},\bm{\theta}))], &\forall \bm{x}\in \{\bm{x}\in \mathcal{X}\setminus \mathcal{T}\mid h(\bm{x})\geq 0\},
    \end{cases}
\end{equation}
then $\mathbb{P}_{\pi}(RA_{\bm{x}_0})\geq p$, $\forall \bm{x}_0\in \mathcal{X}_0$. Moreover, if 
$V: \Omega \to \mathbb{R}$ and $\epsilon>0$ satisfy 
\eqref{second_barrier}, then 
$h(\mathbf{x}) = 1 - (1-p)V(\mathbf{x})$
together with any 
$\lambda \ge \frac{1}{1 + (1-p)\epsilon}$ 
satisfies \eqref{third0_barrier}. Similarly, if a $(\gamma,\delta,\lambda')$-MRAS $V$ satisfies \eqref{mul_barrier}, then $h(\mathbf{x}) = 1 - (1-p)V(\mathbf{x})$
and any $\lambda \in \left[
\max_{V\in[\delta,\lambda']} 
\frac{1 - \frac{1}{\lambda'}V}{1 - \frac{\gamma}{\lambda'}V},
\, 1 \right)$
satisfy \eqref{third0_barrier} with $p = 1 - \frac{1}{\lambda'}$.
\end{corollary}

    \textbf{Computational Tractability}: Solving  \eqref{third_barrier} requires finding both $h$ and $\lambda \in (0,1)$, which is difficult due to the nonlinear term $\lambda \mathbb{E}_{\bm{\theta}}[h(\bm{f}(\bm{x},\bm{\theta}))]$, making it as challenging as solving \eqref{second_barrier}. However, \eqref{third_barrier} is convex in $h(\bm{x})$ if $\lambda$ is fixed. This allows a simpler approach: first, choose a value for $\lambda$ (preferably close to 1 \cite{xue2024sufficient}), then solve for $h(\bm{x})$. If no solution is found, increase $\lambda$ and repeat.

\subsection{Barrier-like Conditions IV}
In this section, we present the fifth barrier-like condition, first proposed in \cite{xue2021reach} to compute $p$-reach-avoid sets in stochastic discrete-time systems. Unlike Bellman-based approaches \cite{xue2024sufficient}, this condition is derived from a different functional equation which gives the exact reach-avoid probability. It has been extended to continuous-time systems (DDEs \cite{xue2021reachd}, ODEs \cite{xue2023reacho,xue2024reach}, SDEs \cite{xue2024}) and to bounding safety probabilities in stochastic discrete-time systems \cite{yu2023safe}.

Like Theorem \ref{first}, this condition uses two functions, $h_1$ and $h_2$, with the requirements: (1) $h_1 \ge p$ on the initial set $\mathcal{X}_0$, (2) $h_1 \le 0$ outside the safe set, (3) $h_1 \le 1$ on the target set $\mathcal{T}$, (4) the expected value of $h_1$ at the next step is at least its current value for $\bm{x} \in \mathcal{X} \setminus \mathcal{T}$, (5) $h_2$ is bounded, and (6) the expected value of $h_2$ increases by at least $h_1(\bm{x})$ at the next step for $\bm{x} \in \mathcal{X} \setminus \mathcal{T}$. If such $h_1$ and $h_2$ exist, the reach-avoid probability from any $\bm{x}_0 \in \mathcal{X}_0$ is at least $p$.

\begin{theorem}[\cite{xue2021reach}]
\label{fifith}
Suppose that there exist a function $h_1 : \mathbb{R}^n \rightarrow \mathbb{R}$ and a bounded function $h_2 : \mathbb{R}^n \rightarrow \mathbb{R}$ satisfying 
\begin{equation}
\tag{BC5}
\label{fifth_barrier}
    \begin{cases}
        h_1(\bm{x})\geq p, &\forall \bm{x}\in \mathcal{X}_0,\\
        h_1(\bm{x}) \leq 0, &\forall \bm{x}\in \mathbb{R}^n\setminus \mathcal{X},\\
        h_1(\bm{x}) \leq 1, & \forall \bm{x}\in \mathcal{T},\\
        \mathbb{E}_{\theta}[h_1(\bm{f}(\bm{x},\bm{\theta}))] \geq h_1(\bm{x}), &\forall \bm{x}\in \mathcal{X}\setminus \mathcal{T},\\
        h_1(\bm{x}) \leq \mathbb{E}_{\theta}[h_2(\bm{f}(\bm{x},\bm{\theta}))]-h_2(\bm{x}), &\forall \bm{x}\in \mathcal{X}\setminus \mathcal{T},
    \end{cases}
\end{equation}
then, for every $\bm{x}_0\in \mathcal{X}_0$, $\mathbb{P}_{\pi}(RA_{\bm{x}_0})\geq p$.
\end{theorem}

\textbf{Theoretical Analysis:} The main difference between \eqref{fifth_barrier} and (11) in \cite{xue2021reach} is that \cite{xue2021reach} uses the set $\widehat{\mathcal{X}}$, which contains all states reachable within one step from $\mathcal{X}$ (i.e., $\mathcal{X}\cup \{\bm{y}\mid \bm{y}=\bm{f}(\bm{x},\bm{\theta}), \bm{x}\in \mathcal{X}, \bm{\theta}\in \Theta\} \subseteq \widehat{\mathcal{X}}$), instead of $\mathbb{R}^n$. The set $\widehat{\mathcal{X}}$ helps define a stopped process by freezing the system \eqref{system} once it leaves the safe set (it must enter $\widehat{\mathcal{X}}$). This also helps remove the invariant assumption in \eqref{second_barrier} and \eqref{mul_barrier}, making the expectations well posed by defining $V: \widehat{\mathcal{X}} \rightarrow \mathbb{R}$. More details on $\widehat{\mathcal{X}}$ can be found in \cite{xue2021reach,xue2024finite}. It can also replace $\mathbb{R}^n$ in \eqref{first_barrier}, \eqref{as_barrier}, and \eqref{third_barrier}. In addition, although \eqref{fifth_barrier} involves two functions, $h_1$ and $h_2$, it is stronger in reach-avoid verification than \eqref{first_barrier}. To clarify, we define $h'_1(\bm{x}) := 1 - h_1(\bm{x})$ and $h'_2(\bm{x}) := 1 - h_2(\bm{x})$, where $h_1$ and $h_2$ satisfy \eqref{fifth_barrier}. Then, we have
\begin{equation}
\label{fifth_trans}
    \begin{cases}
       h'_1(\bm{x})\leq 1-p, &\forall \bm{x}\in \mathcal{X}_0,\\
        h'_1(\bm{x}) \geq 1, &\forall \bm{x}\in \mathbb{R}^n\setminus \mathcal{X},\\
        h'_1(\bm{x}) \geq 0, & \forall \bm{x}\in \mathcal{T},\\
        \mathbb{E}_{\bm{\theta}}[h'_1(\bm{f}(\bm{x},\bm{\theta}))] \leq h'_1(\bm{x}), &\forall \bm{x}\in \mathcal{X}\setminus \mathcal{T},\\
        \mathbb{E}_{\bm{\theta}}[h'_2(\bm{f}(\bm{x},\bm{\theta}))]-h'_2(\bm{x}) \leq  h'_1(\bm{x})-1 , &\forall \bm{x}\in \mathcal{X}\setminus \mathcal{T}.  
    \end{cases}
\end{equation}
Compared to \eqref{first_barrier}, \eqref{fifth_trans} differs in two ways. First, $h_1'(\bm{x})$ need not be non-negative over $\mathcal{X} \setminus \mathcal{T}$. Second, the expected value of $h'_2(\bm{x})$ at the next step is not required to decrease by a fixed positive amount for all $\bm{x} \in \mathcal{X} \setminus \mathcal{T}$. These relaxations make \eqref{fifth_barrier} applicable to systems where all trajectories may leave $\mathcal{X}$ eventually. 
On the other hand,  condition~\eqref{fifth_barrier} has been shown to be necessary for reach-avoid verification under certain assumptions. Corollary 1 in \cite{xue2024sufficient} implies that when the initial set $\mathcal{X}_0$ is a singleton ${\bm{x}_0}$, if the verified threshold $p$ is strictly less than $\mathbb{P}_{\pi}(RA_{\bm{x}_0})$, 
 then there definitely exist a function $h_1: \mathbb{R}^n \rightarrow \mathbb{R}$ and a  bounded function $h_2: \mathbb{R}^n \rightarrow \mathbb{R}$ such that \eqref{fifth_barrier} holds with $h_1(\bm{x}_0)\geq p$. 
Additionally, reversing the inequality in \eqref{fifth_barrier} gives a condition whose feasibility provides an upper bound on the reach-avoid probability. Corollary \ref{barrier_upper} follows directly from the proof of Corollary 2 in \cite{xue2021reach}, with the inequality reversed.

\begin{corollary}
\label{barrier_upper}
    Suppose that there exist a function $h_1 : \mathbb{R}^n \rightarrow \mathbb{R}$ and a bounded function $h_2 : \mathbb{R}^n \rightarrow \mathbb{R}$ satisfying 
\begin{equation}
\label{fifth_barrier_upp}
    \begin{cases}
        h_1(\bm{x})\leq p, &\forall \bm{x}\in \mathcal{X}_0,\\
        h_1(\bm{x}) \geq 0, &\forall \bm{x}\in \mathbb{R}^n\setminus \mathcal{X},\\
        h_1(\bm{x}) \geq 1, & \forall \bm{x}\in \mathcal{T},\\
        \mathbb{E}_{\theta}[h_1(\bm{f}(\bm{x},\bm{\theta}))] \leq h_1(\bm{x}), &\forall \bm{x}\in \mathcal{X}\setminus \mathcal{T},\\
        h_1(\bm{x}) \geq \mathbb{E}_{\theta}[h_2(\bm{f}(\bm{x},\bm{\theta}))]-h_2(\bm{x}), &\forall \bm{x}\in \mathcal{X}\setminus \mathcal{T},
    \end{cases}
\end{equation}
then, for every $\bm{x}_0\in \mathcal{X}_0$, $\mathbb{P}_{\pi}(RA_{\bm{x}_0})\leq p$.
\end{corollary}

    \textbf{Computational Tractability}: Similar to the condition \eqref{first_barrier}, the condition \eqref{fifth_barrier} is also convex in $h_1(\bm{x})$ and $h_2(\bm{x})$. 
    This convexity property provides several theoretical and computational benefits, as previously discussed.  


\textbf{SUMMARY on Comparisons above:} In general, all of the above conditions are sound but not complete, although conditions \eqref{third_barrier} and \eqref{fifth_barrier} can achieve completeness under certain assumptions. Condition~\eqref{first_barrier} is very restrictive: it requires a positive probability of remaining safe forever and zero probability of staying indefinitely in $\mathcal{X}\setminus \mathcal{T}$. This is too strong for many practical systems where trajectories may leave the safe set after completing their task. Whenever condition~\eqref{first_barrier} holds, condition~\eqref{as_barrier} can also be used for reach-avoid verification, and in this case, it is both necessary and sufficient.
 Conditions \eqref{second_barrier} and \eqref{mul_barrier} are limited by their reliance on a pre-defined robust invariant set $\Omega$, which is hard to find or may not exist if $\Omega \neq \mathbb{R}^n$. In contrast, \eqref{third_barrier} and \eqref{fifth_barrier} avoid both the infinite-horizon safety requirement and the need for a robust invariant, making them applicable to a broader class of systems. Under assumptions such as $\mathcal{X}_0$ being a singleton ${\bm{x}_0}$ and the threshold $p$ being strictly less than the exact reach-avoid probability $\mathbb{P}_{\pi}(RA_{\bm{x}_0})$, conditions \eqref{third_barrier} and \eqref{fifth_barrier} are also necessary (i.e., complete).
We further relate \eqref{third_barrier} to \eqref{second_barrier} and \eqref{mul_barrier} via condition \eqref{third0_barrier} in Corollary \ref{third_cor}, by requiring $h(\bm{x}) \leq \lambda \mathbb{E}_{\bm{\theta}}[h(\bm{f}(\bm{x},\bm{\theta}))]$ to hold over $\{\bm{x} \in \mathcal{X} \setminus \mathcal{T} \mid h(\bm{x}) \geq 0\}$ rather than the full set $\mathcal{X} \setminus \mathcal{T}$. This shows that the feasible set of \eqref{third0_barrier} is at least as large as that of \eqref{second_barrier} and \eqref{mul_barrier}, potentially making it more powerful for reach-avoid verification than \eqref{second_barrier} and \eqref{mul_barrier}.

Finally, \eqref{first_barrier} and \eqref{fifth_barrier} are convex in the unknown functions, which aids practical computation. In contrast, \eqref{second_barrier}, \eqref{mul_barrier}, and \eqref{third_barrier} are nonlinear, though the nonlinearity of \eqref{third_barrier} can be removed by fixing $\lambda$, making it convex.

\section{Examples and Discussions}
\label{sec:ex}
This section presents numerical experiments on  barrier-like conditions for reach–avoid verification, using polynomial and neural network barriers with common computational methods-SDP and CEGIS. Conditions \eqref{first_barrier}, \eqref{third_barrier}, and \eqref{fifth_barrier} are convex in the barrier functions, allowing polynomial barriers to be computed via SDPs when $\bm{f}(\bm{x},\bm{\theta})$ is polynomial in $\bm{x}$ and the sets $\mathcal{X}_0$, $\mathcal{T}$, and $\mathcal{X}$ are semi-algebraic. They are encoded as semidefinite constraints in YALMIP~\cite{lofberg2004yalmip} and solved with Mosek, restricting polynomial coefficients to $[-100,100]$ for numerical stability. Neural network barriers, in contrast, handle both polynomial and general nonlinear systems using CEGIS. The neural network is iteratively refined by alternating between (i) synthesis, optimizing network parameters on a finite sample set, and (ii) verification, finding counterexamples via Lipschitz-based discretization~\cite{badings2025policy,shakhesi2025counterexample,vzikelic2023learning}. Counterexamples are added to the training set, and the process repeats until no violations remain. In the examples, we use neural networks with $4\times4$ and $8\times 8$ hidden layers and limit the number of iterations to 10 for demonstration. Moreover,
we follow \cite{badings2025policy,vzikelic2023learning} and restrict the computation to a compact set such as $\Omega$ or the one-step over-approximated reachable set $\widehat{\mathcal{X}}$.





\subsection{Examples}

\begin{example}
\label{ex1} 
We consider an example from \cite{xue2021reach}:
\begin{equation*}
\begin{cases}
x(l+1)=x(l)+0.01(-0.5x(l)-0.5y(l)+0.5x(l)y(l)),\\
y(l+1)=y(l)+0.01(-0.5y(l)+1+\theta(l)), 
\end{cases}
\end{equation*}
where  $\theta : [0,\infty)\rightarrow \Theta$ with $\Theta$ being a set in $\mathbb{R}$. 

Consider $\mathcal{X}_0=\{(x,y)^{\top} \mid (x+0.2)^2+(y+0.8)^2\leq 10^{-4}\}$, $\mathcal{X}=\{(x,y)^{\top} \mid x^2+y^2\leq 1\}$, and $\mathcal{T}=\{(x,y)^{\top} \mid 10x^2+10(y-0.5)^2\leq 1\}$. The disturbance $\theta(l)$ is uniformly distributed in $\Theta=[-10,10]$. The SDP feasibility of barrier-like conditions \eqref{first_barrier}, \eqref{third_barrier}, and \eqref{fifth_barrier} under different probability thresholds $p$ is summarized in Table~\ref{tab:sdp_ex1}. In contrast, the neural implementation based on the CEGIS framework failed to synthesize valid barrier-like functions. In this example, a robust invariant set may not exist; hence, only conditions~\eqref{first_barrier}, \eqref{third_barrier}, and \eqref{fifth_barrier} are applied. Furthermore, since the assumption in Proposition~\ref{as} cannot be guaranteed, condition~\eqref{as_barrier} is not applied. In these conditions, a one-step reachable set  $\widehat{\mathcal{X}} = [-1.015,1.015] \times [-1.085,1.105]$ is used for computations rather than $\mathbb{R}^n$. For $\widehat{\mathcal{X}}$, please refer to the discussion following Theorem~\ref{fifith}. 




        
\end{example}

\vspace{-10pt} 
\begin{table}[ht]
\caption{\centering SDP Feasibility and Computation Times (in seconds) for Example~\ref{ex1}\\
(\XSolidBrush: infeasible; 
\CheckmarkBold: feasible)}
\label{tab:sdp_ex1}
\centering
\begin{tabular}{|c|c|c|c|c|}
\hline
Degree & $p$ & \eqref{first_barrier}  with $\epsilon=10^{-6}$ & \eqref{third_barrier} with $\lambda=0.9999$ &\eqref{fifth_barrier}\\
\hline
8 & 0.15& \XSolidBrush ($T=2.40$) &\CheckmarkBold($T=2.45$) & \CheckmarkBold($T=2.61$)  \\
10 & 0.50&\XSolidBrush ($T=5.92$) & \CheckmarkBold($T=5.20$) & \CheckmarkBold($T=5.55$) \\
16 &0.65&\XSolidBrush ($T=17.45$) & \CheckmarkBold($T=17.96$) & \CheckmarkBold($T=18.75$) \\
\hline
\end{tabular}
\end{table}
\begin{table}[ht]
\caption{\centering SDP Feasibility and Computation Times $T$ (in seconds) for Example~\ref{ex2}\\
(\XSolidBrush: infeasible; 
\CheckmarkBold: feasible)}
\label{tab:sdp_ex2}
\centering
\begin{tabular}{|c|c|c|c|c|c|}
\hline
Degree & $p$ & \eqref{first_barrier} with $\epsilon=10^{-6}$& \eqref{as_barrier} & \eqref{third_barrier} with $\lambda=9\times 10^{-4}$ &\eqref{fifth_barrier}\\
\hline
8 & 0.60& \XSolidBrush($T=2.21$) &\CheckmarkBold($T=1.95$) &\CheckmarkBold($T=1.66$) & \CheckmarkBold($T=1.83$)  \\
10 & 0.85&\CheckmarkBold($T=2.83$) &\CheckmarkBold($T=2.48$) &\CheckmarkBold($T=2.50$) & \CheckmarkBold($T=2.95$) \\
12 &0.90 &\CheckmarkBold($T=5.84$) & \CheckmarkBold($T=5.19$)&\CheckmarkBold($T=5.20$) & \CheckmarkBold($T=6.13$) \\
\hline
\end{tabular}
\end{table}
\begin{table}[h!]
\caption{\centering SDP Feasibility and Computation Times $T$ (in seconds) for Example~\ref{ex3}\\
(\XSolidBrush: infeasible; 
\CheckmarkBold: feasible)}
\label{tab:sdp_ex3}
\centering
\begin{tabular}{|c|c|c|c|c|c|}
\hline
Degree & $p$ & \eqref{first_barrier} with $\epsilon=10^{-6}$& \eqref{as_barrier} & \eqref{third_barrier} with $\lambda=9\times 10^{-4}$ &\eqref{fifth_barrier}\\
\hline
4 & 0.80& \CheckmarkBold($T=0.72$) &\CheckmarkBold($T=0.42$) &\CheckmarkBold($T=0.49$) & \CheckmarkBold($T=0.75$)  \\
\hline
\end{tabular}
\end{table}
\vspace{-0pt}

\begin{example}
\label{ex2}
Consider the following discrete-time Lotka-Volterra model from \cite{xue2021reach}:
\begin{equation}
\begin{cases}
x(l+1)=rx(l)-ay(l)x(l),\\
y(l+1)=sy(l)+acy(l)x(l),
\end{cases}
\end{equation}
where $r=0.5$, $a=1$, $s=-0.5+\theta(l)$ with $\theta : \mathbb{N}\rightarrow [-1,1]$ and $c=1$.

Assume the random vector $\theta(l)$, $l\in\mathbb{N}$, is uniformly distributed on $[-1,1]$. Let $\mathcal{X}_0 = \{(x,y)^\top \mid (x+0.6)^2 + (y+0.5)^2 \le 10^{-4}\}$, $\mathcal{X} = \{(x,y)^\top \mid x^2 + y^2 - 4 \le 0\}$, and $\mathcal{T} = \{(x,y)^\top \mid 100x^2 + 100y^2 \le 1\}$. The SDP feasibility of the barrier-like conditions \eqref{first_barrier}, \eqref{third_barrier}, and \eqref{fifth_barrier_upp} for different probability thresholds $p$ is summarized in Table \ref{tab:sdp_ex2}. Condition~\eqref{first_barrier} successfully verifies the reach-avoid specification for $p=0.85$ and $0.90$, confirming the assumption in Proposition~\ref{as}. Thus, condition \eqref{as_barrier} is also applied.
Since this example may not have a robust invariant set, only conditions \eqref{first_barrier}, \eqref{as_barrier}, \eqref{third_barrier}, and \eqref{fifth_barrier} are used. Computations are performed over the one-step reachable set $\widehat{\mathcal{X}} = [-5,5]\times[-7,7]$ instead of $\mathbb{R}^n$ (see discussion after Theorem \ref{fifith}). In contrast, the CEGIS method failed to synthesize valid barrier-like functions when using conditions \eqref{first_barrier}, \eqref{as_barrier}, \eqref{third_barrier}, and \eqref{fifth_barrier}.

\end{example}




\begin{example}
\label{ex3}
Consider the following reach-avoid problem from \cite{vzikelic2023learning}:
\begin{equation*}
\begin{cases}
x(l+1)=0.6 x(l) +  0.05 y(l) + 0.01 \theta_1 (l),\\
y(l+1)=0.6 y(l)+ 0.005 \theta_2 (l),
\end{cases}
\end{equation*}
where $\theta_1(l)$ and $\theta_2(l)$ are independent and follow a triangular distribution on $[-1,1]$ with density Triangular$(\theta) = 1 - |\theta|$ if $|\theta| < 1$, otherwise $0$.

The initial set is $\mathcal{X}_0 = ([-0.15,-0.1]\cup[0.1,0.15])\times[-0.1,0.1]$, the target set is $\mathcal{T} = [-0.1,0.1]^2$, the safe set is $\mathcal{X} = [-0.6,0.6]^2$, and the invariant set is $\Omega = [-1,1]^2$. Since this example has a robust invariant set $\Omega$, all conditions~\eqref{first_barrier}, \eqref{second_barrier}, \eqref{mul_barrier}, \eqref{third_barrier}, \eqref{third0_barrier}, and \eqref{fifth_barrier} are used for reach-avoid verification. SDP results are summarized in Table~\ref{tab:sdp_ex3}, and CEGIS results are in Table~\ref{ex3_tab1}.
\end{example}

\vspace{-35pt}
\begin{table}[!h]
\centering
\caption{\centering Feasibility and Computation Times $T$ (in seconds)for Example \ref{ex3} with $p=0.6$ and $p=0.8$(\XSolidBrush: infeasible; \CheckmarkBold: feasible)}
\label{ex3_tab1}
\begin{tabular}{cccccc}
\toprule
\multirow{2}{*}{\textbf{Condition}} &\multirow{2}{*}{\textbf{Hidden layers}} & \multicolumn{2}{c}{\textbf{$p=0.6$}} & \multicolumn{2}{c}{\textbf{$p=0.8$}} \\ 
\cmidrule(lr){3-4} \cmidrule(lr){5-6}
 && \textbf{Time} & \textbf{Feasibility} & \textbf{Time} & \textbf{Feasibility}  \\ 
\midrule
\eqref{first_barrier} & 4$\times$4 & 367.30& \XSolidBrush &360.51 & \XSolidBrush   \\ 
\eqref{as_barrier} &4$\times$4 & 11.67 & \CheckmarkBold &110.13&\XSolidBrush\\
\eqref{second_barrier} & 4$\times$4 & 14.83& \CheckmarkBold & 152.7& \XSolidBrush \\ 
\eqref{mul_barrier} & 4$\times$4 & \textbf{6.60}&\CheckmarkBold  & 61.05&\XSolidBrush  \\ 
\eqref{third_barrier} & 4$\times$4 &21.05 &\CheckmarkBold&\textbf{21.86}&\CheckmarkBold\\\
\eqref{third0_barrier} & 4$\times$4 & 17.11&\CheckmarkBold&13.80 &\CheckmarkBold\\
\eqref{fifth_barrier} & $(h_1: 4\times4, h_2: 8\times8)$ &34.02&\CheckmarkBold&33.30&\CheckmarkBold\\
\cmidrule(lr){1-6}
\eqref{first_barrier} & 8$\times$8& 657.53& \XSolidBrush & 661.91& \XSolidBrush\\
\eqref{as_barrier} & 8$\times$8&18.81&\CheckmarkBold &32.61  &\CheckmarkBold\\
\eqref{second_barrier} & 8$\times$8 &16.56 & \CheckmarkBold &31.11&\CheckmarkBold\\
\eqref{mul_barrier} & 8$\times$8 & \textbf{7.18}& \CheckmarkBold&\textbf{14.05}&\CheckmarkBold\\
\eqref{third_barrier} & 8$\times$8 & 19.24&\CheckmarkBold&33.44&\CheckmarkBold\\
\eqref{third0_barrier} & 8$\times$8 &12.81&\CheckmarkBold&28.71&\CheckmarkBold\\
\eqref{fifth_barrier} & $(h_1: 8\times8, h_2: 8\times8)$ &42.79 &\CheckmarkBold&64.92&\CheckmarkBold\\
\bottomrule
\end{tabular}
\end{table}
\vspace{-40pt}
\begin{table}[H]
\centering
\caption{\centering Feasibility and Computation Times $T$ (in seconds)for Example \ref{ex4} with $p=0.4$ and $p=0.6$(\XSolidBrush: infeasible; \CheckmarkBold: feasible)}
\label{ex4_tab1}
\begin{tabular}{cccccc}
\toprule
\multirow{2}{*}{\textbf{Condition}} &\multirow{2}{*}{\textbf{Hidden layers}} & \multicolumn{2}{c}{\textbf{$p=0.4$}} & \multicolumn{2}{c}{\textbf{$p=0.6$}} \\ 
\cmidrule(lr){3-4} \cmidrule(lr){5-6}
 && \textbf{Time} & \textbf{Feasibility} & \textbf{Time} & \textbf{Feasibility}  \\ 
\midrule
\eqref{first_barrier} & 4$\times$4 & 154.11& \XSolidBrush &151.86 & \XSolidBrush   \\ 
\eqref{third_barrier} & 4$\times$4 &\textbf{82.90}&\CheckmarkBold&\textbf{173.61}&\CheckmarkBold\\\
\eqref{fifth_barrier} & $(h_1: 4\times4, h_2: 8\times8)$ &143.91&\CheckmarkBold&183.31&\CheckmarkBold\\
\cmidrule(lr){1-6}
\eqref{first_barrier} & 8$\times$8& 188.80& \XSolidBrush & 205.94& \XSolidBrush\\
\eqref{third_barrier} & 8$\times$8 & \textbf{60.51}&\CheckmarkBold&172.35&\CheckmarkBold\\
\eqref{fifth_barrier} & $(h_1: 8\times8, h_2: 8\times8)$ &116.31&\CheckmarkBold&\textbf{47.47}&\CheckmarkBold\\
\bottomrule
\end{tabular}
\end{table}
\vspace{-20pt}

\begin{example}
\label{ex4}
We consider the feedback control of nonlinear system :
\begin{equation*}
\begin{cases}
x(l+1)=x(l)+0.2y(l),\\
y(l+1)=y(l)+0.2(\sin x(l)-y(l)+u(l)+\theta(l)), 
\end{cases}
\end{equation*}
where  $u(l) = -3x(l)-\frac{1}{2}y(l)$ and  $\theta : \mathbb{N}\rightarrow [-0.05,0.05]$.

Assume the random $\theta(l)$, $l \in \mathbb{N}$, is uniformly distributed on $[-0.05,0.05]$, $\mathcal{X}_0 = [0.2,0.3]\times[-0.1,0.1]$, $\mathcal{X} = [-1,1]^2$, and $\mathcal{T} = [-0.2,0.2]^2$. As before, computations use the one-step reachable set $\widehat{\mathcal{X}} = [-1.2,1.2]\times[-1.51,1.51]$ instead of $\mathbb{R}^n$. Since the system is nonlinear and not polynomial, SDP cannot be applied, but CEGIS can synthesize neural barriers. Results are shown in Tables~\ref{ex4_tab1}.


\end{example}


\subsection{Discussions}

From the numerical experiments, we observe that all the reach-avoid verification problems can be solved using \eqref{third_barrier} and \eqref{fifth_barrier}, with either the SDP or CEGIS approach. 
\eqref{as_barrier} outperforms \eqref{first_barrier}(e.g., Examples \ref{ex2} and \ref{ex3}), and \eqref{third0_barrier} outperforms  \eqref{second_barrier} and \eqref{mul_barrier} for systems (e.g., Example \ref{ex3}).

For Examples~\ref{ex1}–\ref{ex3}, which involve polynomial systems, both SDP and CEGIS can be to employed to solve the reach-avoid problems. The experiment results show that SDP outperforms than CEGIS for these polynomial examples. The SDP method encodes the barrier conditions as a single convex optimization problem, making it straightforward to implement and generally efficient. However, SDP applies only to polynomial systems, and its complexity grows quickly with system size and polynomial degree. 
 CEGIS exhibits more variable behavior, and no single condition consistently performs best—especially in terms of computation time. This variability arises because CEGIS is heuristic and highly sensitive to choices such as the neural network architecture and counterexamples. For Examples \ref{ex1} and \ref{ex2}, CEGIS failed to find feasible solutions but SDP works well, and it remains unclear how to tune the network parameters to ensure success. Additionally, the different structures of the barrier conditions make fair comparisons using CEGIS difficult. Intuitively, conditions \eqref{second_barrier}, \eqref{mul_barrier}, and \eqref{third0_barrier} compute expectations over smaller domains, which can reduce computation time. In contrast, \eqref{third_barrier} evaluates expectations over a larger region, potentially increasing verification time. \eqref{fifth_barrier} provides stronger theoretical guarantees but uses two coupled functions, adding computational cost. However, these intuitions do not always match practical results. Moreover, CEGIS is iterative but provides no guarantee of termination, and counterexamples do not always improve the candidate barrier, which limits its broader applicability. On the other hand, neural barrier methods can handle non-polynomial systems and are more scalable to train, but formally verifying them efficiently remains challenging. Existing verification techniques such as SMT solvers \cite{zhao2020synthesizing,abate2021fossil}, mixed-integer programming  \cite{zhao2022verifying}, and Lipschitz-based discretization \cite{badings2025policy,shakhesi2025counterexample} face severe scalability issues.  The above discussions highlight the need for more advanced methods for handling these barrier-like conditions.

\section{Conclusion}
\label{sec:con}
This paper compared several representative barrier-like conditions from the literature for infinite-horizon reach-avoid verification in stochastic discrete-time systems. Our comparisons, conducted from both theoretical and computational perspectives, highlight their relative strengths and limitations, providing valuable insights for practical applications.
\bibliographystyle{unsrt}
\bibliography{reference}
\newpage
\section{Appendix}

\textbf{The proof of Corollary \ref{first_coro}}:

\begin{proof}
Let $\{\mathcal{F}_l\}_{l\ge 0}$ be the natural filtration generated by the process:
\[
\mathcal{F}_l := \sigma\big(\bm{\phi}_\pi^{\bm{x}_0}(0),\dots,\bm{\phi}_\pi^{\bm{x}_0}(l)\big).
\]

\textbf{Part 1: Safety guarantee via $h_1$.}  
From the non-negative supermartingale property of $h_1$, Ville's inequality implies
$\mathbb{P}_\pi\Big(\forall i \in \mathbb{N}. \; \bm{\phi}_\pi^{\bm{x}_0}(i) \in \mathcal{X}\Big) \ge 1 - h_1(\bm{x}_0)$.

\textbf{Part 2: Infinite visits to $\mathcal{T}$ conditional on safety.}
Assume $p > 0$ and $h_2$ is bounded. Define the event of finite visits while staying in $\mathcal{X}$:
\[
A := \left\{ \pi \mid \forall i \ge 0. \bm{\phi}_\pi^{\bm{x}_0}(i) \in \mathcal{X} \text{ and } \sum_{i=0}^\infty 1_{\mathcal{T}}(\bm{\phi}_\pi^{\bm{x}_0}(i)) < \infty \right\}.
\]

For each $M \in \mathbb{N}$, define the event of exactly $M$ visits:
\[
A_M := \left\{ \pi\mid \forall i \ge 0. \bm{\phi}_\pi^{\bm{x}_0}(i) \in \mathcal{X} \text{ and } \sum_{i=0}^\infty 1_{\mathcal{T}}(\bm{\phi}_\pi^{\bm{x}_0}(i)) = M \right\}.
\]
Then $A = \bigcup_{M \in \mathbb{N}} A_M$.

For each $M \in \mathbb{N}$, define the stopping time:
\[
\tau_M := \inf \left\{ k \ge 0: \sum_{i=0}^k 1_{\mathcal{T}}(\bm{\phi}_\pi^{\bm{x}_0}(i)) = M \right\},
\]
with $\inf \emptyset = \infty$. This is a valid stopping time since $\{\tau_M \le l\} \in \mathcal{F}_l$ for all $l$. Moreover, define $\nu:=\inf\{k>0: \bm{\phi}_\pi^{\bm{x}_0}(\tau_M+k)\notin \mathcal{X}\setminus \mathcal{T}\}$. On the event above, $\nu=\infty$.


Define the stopped process: $Y_m(\pi) := h_2\big(\bm{\phi}_\pi^{\bm{x}_0}(\tau_M + \min\{m,\nu\})\big), m \ge 0$, with the filtration $\mathcal{G}_m:= \mathcal{F}_{\tau_M + \min\{m,\nu\}}$. Assume $0\le h_2(\bm{x}) \leq B$ over $\mathbb{R}^n$. We can conclude that $0\leq Y_m(\pi) \leq B$ and thus $\{Y_m(\pi)\}$ is bounded and integrable.

Whenever $1\leq k<\nu$, the state $\bm{\phi}_\pi^{\bm{x}_0}(\tau_M+k)\in \mathcal{X}\setminus \mathcal{T}$. Thus, by the $\epsilon$-ranking supermartingale property of $h_2$, for each $k \ge 0$:
\[
\mathbb{E}_\pi[Y_{k+1} \mid \mathcal{G}_k] \le Y_k - \epsilon.
\]

By induction, for any $m \ge 0$, we have
$\mathbb{E}_\pi[Y_m] \leq B -  \epsilon \mathbb{E}_{\pi}[\max\{0,\min\{m,\nu\}\}-1]$. If $\mathbb{P}_{\pi}(\nu=\infty)>0$, then $\mathbb{E}_{\pi}[\min\{m,\nu\}]\geq m \mathbb{P}_{\pi}(\nu=\infty)$, so the right-hand side tends to $-\infty$ as $m\rightarrow \infty$, a contradiction. Hence, $\mathbb{P}_{\pi}(\nu=\infty)=0$, ruling out the possibility of staying in $\mathcal{X}\setminus \mathcal{T}$ forever. 



Therefore, conditioned on remaining in $\mathcal{X}$, we obtain
\[
\mathbb{P}_\pi\left( \sum_{i=1}^\infty 1_{\mathcal{T}}(\bm{\phi}_\pi^{\bm{x}_0}(i)) = \infty \;\middle|\; \forall i \in \mathbb{N}, \bm{\phi}_\pi^{\bm{x}_0}(i) \in \mathcal{X} \right) = 1.
\]
The conclusion is proved.
\end{proof}

\textbf{The proof of Corollary \ref{third_cor}}

\begin{proof}
1) Firstly, we show $\mathbb{P}_{\pi}(RA_{\bm{x}_0})\geq p, \forall \bm{x}_0\in \mathcal{X}_0$.

Since $h(\bm{x})\leq 0, \forall \bm{x}\in \Omega\setminus \mathcal{X}$, $\{\bm{x}\in \Omega \mid h(\bm{x})>0\}$ is a subset of $\mathcal{X}$, i.e., $\{\bm{x}\in \Omega \mid h(\bm{x})>0\}\subseteq \mathcal{X}$. Obviously, $\mathcal{X}_0\subseteq \{\bm{x}\in \Omega \mid h(\bm{x})>0\}$.


Let $A=\{\bm{x}\in \mathcal{X}\mid h(\bm{x})\geq  0\}\cap \mathcal{T}$, $B=\{\bm{x}\in \mathcal{X}\setminus \mathcal{T}\mid h(\bm{x})\geq 0\}$, and $C=\{\bm{x}\in \mathcal{X}\mid h(\bm{x})\geq 0\}$. Obviously, $A\cup B=C$. 
We consider the stopped process $\{\tilde{\bm{\phi}}_{\pi}^{\bm{x}_0}(k)\}_{k\geq 0}$ satisfying 
\[\tilde{\bm{\phi}}_{\pi}^{\bm{x}_0}(k+1)=\tilde{\bm{f}}(\tilde{\bm{\phi}}_{\pi}^{\bm{x}_0}(k),\bm{\theta}(k)),\]
where $\tilde{\bm{f}}(\bm{x},\bm{\theta}):=1_{A \cup (\Omega\setminus C) }(\bm{x})\cdot \bm{x} +1_{B}(\bm{x})\cdot\bm{f}(\bm{x},\bm{\theta})$.

Thus, the reach-avoid probability $\mathbb{P}_{\pi}(RA_{\bm{x}_0})$ for $\bm{x}_0\in \mathcal{X}_0$ is larger than or equal to the probability $\mathbb{P}_{\pi}(R_{\bm{x}_0})$ of reaching the  set $A$ eventually for the stochastic process $\{\tilde{\bm{\phi}}_{\pi}^{\bm{x}_0}(k)\}_{k\geq 0}$, where $R_{\bm{x}_0}=\{\pi\mid \exists k\in \mathbb{N}.\tilde{\bm{\phi}}_{\pi}^{\bm{x}_0}(k)\in A\}$. Since $A_i\subseteq A_j$ for $i\leq j$, where $A_i=\{\pi\mid \tilde{\bm{\phi}}_{\pi}^{\bm{x}_0}(i)\in A\}$ and $A_j=\{\pi\mid \tilde{\bm{\phi}}_{\pi}^{\bm{x}_0}(j)\in A\}$, we have $\mathbb{P}_{\pi}(A_i) \leq \mathbb{P}_{\pi}(A_j) \text{~and~}\mathbb{P}_{\pi}(R_{\bm{x}_0})=\lim_{i\rightarrow \infty}\mathbb{P}_{\pi}(A_i)$.

On the other hand, since $h(\bm{x})\leq 1, \forall \bm{x}\in \Omega$, we have $h(\bm{x})\leq 1, \forall \bm{x}\in \mathcal{T}$, and thus $h(\bm{x})\leq \lambda \mathbb{E}_{\bm{\theta}}[h(\tilde{\bm{f}}(\bm{x},\bm{\theta}))]+(1-\lambda)1_{A}(\bm{x}), \forall \bm{x}\in \Omega$.

Via induction, we have, for $\bm{x}_0\in \mathcal{X}_0$,
\[
\begin{split}
h(\bm{x}_0)&\leq \lambda^i \mathbb{E}_{\pi}[h(\tilde{\bm{\phi}}_{\pi}^{\bm{x}_0}(i))]+(1-\lambda)\frac{1-\lambda^i}{1-\lambda}\mathbb{E}_{\pi}[1_{A}(\tilde{\bm{\phi}}_{\pi}^{\bm{x}_0}(i-1))]\\
&\leq \lambda^i+(1-\lambda^i)\mathbb{E}_{\pi}[1_{A}(\tilde{\bm{\phi}}_{\pi}^{\bm{x}_0}(i-1))],
\end{split}
\]
where $\mathbb{E}_{\pi}[1_{A}(\tilde{\bm{\phi}}_{\pi}^{\bm{x}_0}(i-1))]=\mathbb{P}_{\pi}(A_{i-1})$. 
Let $i\rightarrow \infty$, we have 
\[h(\bm{x}_0)\leq \lim_{i\rightarrow \infty} \mathbb{E}_{\pi}[1_{A}(\tilde{\bm{\phi}}_{\pi}^{\bm{x}_0}(i-1))]=\mathbb{P}_{\pi}(R_{\bm{x}_0})\leq \mathbb{P}_{\pi}(RA_{\bm{x}_0}).\]

Since $h(\bm{x}_0)\geq p, \forall \bm{x}\in \mathcal{X}_0$, we have the conclusion.

2) Secondly, we will show that if $V : \Omega\rightarrow \mathbb{R}$ and $\epsilon>0$ satisfy \eqref{second_barrier}, then $h(\bm{x}):=(1-(1-p)V(\bm{x}))$ and $\lambda\geq \frac{1}{1+(1-p)\epsilon}$ will satisfy \eqref{third0_barrier}. 

Let $h(\bm{x}):=(1-(1-p)V(\bm{x}))$. It is easy to show that 
\begin{equation}
    \begin{cases}
        h(\bm{x}) \leq 0, &\forall \bm{x}\in \Omega\setminus \mathcal{X},\\
        h(\bm{x})\geq p,&\forall \bm{x} \in \mathcal{X}_0.
    \end{cases}
\end{equation}

Since $V(\bm{x})\geq 0$ for $\bm{x}\in \Omega$, $h(\bm{x})\leq 1$ for $\bm{x}\in \Omega$. In addition, since \[\mathbb{E}_{\bm{\theta}}[V(\bm{f}(\bm{x},\bm{\theta}))]-V(\bm{x})\leq -\epsilon\] holds for $\bm{x}\in \{\Omega\setminus \mathcal{T}\mid V(\bm{x})\leq  \frac{1}{1-p}\}$, we have $\mathbb{E}_{\bm{\theta}}[h(\bm{f}(\bm{x},\bm{\theta}))]-h(\bm{x})\geq \epsilon(1-p)$ for $\bm{x}\in \{\Omega\setminus \mathcal{T}\mid h(\bm{x})\geq 0\}$. 
Also, since $h(\bm{x})\leq 1$ for $\bm{x}\in \Omega$, we have
$\mathbb{E}_{\bm{\theta}}[h(\bm{f}(\bm{x},\bm{\theta}))]-h(\bm{x})\geq \epsilon(1-p) \geq \epsilon (1-p) h(\bm{x})$ for $\bm{x}\in \{\Omega\setminus \mathcal{T}\mid h(\bm{x})\geq 0\}$,  implying
$\mathbb{E}_{\bm{\theta}}[h(\bm{f}(\bm{x},\bm{\theta}))]\geq (1+\epsilon (1-p)) h(\bm{x})$ for $\bm{x}\in \{\Omega\setminus \mathcal{T}\mid h(\bm{x})\geq 0\}$. Therefore, the conclusion holds.

3) Thirdly, we will show that there exists a $(\gamma,\delta,\lambda')$-multiplicative reach-avoid supermartingale $V : \Omega\rightarrow \mathbb{R}$ satisfying \eqref{mul_barrier}, then $h(\bm{x}):=(1-(1-p)V(\bm{x}))$ and any $\lambda\in [\max_{V\in[\delta,\lambda']} \frac{1-\frac{1}{\lambda'}V}{1-\frac{\gamma}{\lambda'}V},1)$ satisfy \eqref{third0_barrier} with $1-\frac{1}{\lambda'}=p$.


Since $V(\bm{x})\leq 1$ for $\bm{x}\in \mathcal{X}_0$, $h(\bm{x})=1-(1-p)V(\bm{x})\geq 1-(1-p)=p, \forall \bm{x}\in \mathcal{X}_0$ holds; from  $V(\bm{x})\geq \lambda' =\frac{1}{1-p}, \forall \bm{x}\in \Omega\setminus \mathcal{X}$, we have $h(\bm{x})\leq 0, \forall \bm{x}\in \Omega\setminus \mathcal{X}$; since $V(\bm{x})\geq 0, \forall \bm{x}\in \Omega$, we have $h(\bm{x}) \leq 1, \forall \bm{x}\in \Omega$. 

Since $\mathbb{E}_{\bm{\theta}}[V(\bm{f}(\bm{x},\bm{\theta}))]\leq \gamma V(\bm{x}), \forall \bm{x}\in \{\Omega\setminus \mathcal{T}\mid V(\bm{x})\leq \lambda'=\frac{1}{1-p}\}$, we have, for $\bm{x}\in \{\Omega\setminus \mathcal{T}\mid h(\bm{x})\geq 0\}$, $1-\frac{1}{\lambda'}\gamma V(\bm{x})>0$ holds and thus,  
\begin{equation*}
\begin{split}
&\lambda \mathbb{E}_{\bm{\theta}}[h(\bm{f}(\bm{x},\bm{\theta}))]-h(\bm{x})
\\
=&\lambda(1-(1-p)\mathbb{E}_{\bm{\theta}}[V(\bm{f}(\bm{x},\bm{\theta}))])-(1-(1-p)V(\bm{x}))\\
=& \lambda(1-\frac{1}{\lambda'}\mathbb{E}_{\bm{\theta}}[V(\bm{f}(\bm{x},\bm{\theta}))])-(1-\frac{1}{\lambda'}V(\bm{x}))\\
\geq &\lambda(1-\frac{1}{\lambda'}\gamma V(\bm{x}))-(1-\frac{1}{\lambda'}V(\bm{x}))\\
\geq &\max_{V\in[\delta,\lambda']} \frac{1-\frac{1}{\lambda'}V}{1-\frac{\gamma}{\lambda'}V}(1-\frac{1}{\lambda'}\gamma V(\bm{x}))-(1-\frac{1}{\lambda'}V(\bm{x}))\geq 0.
\end{split}
\end{equation*}
The conclusion is proved.

\end{proof}

\end{document}